# Nonlinear Pulse Equi-partition in Weakly Coupled Ordered Granular Chains with no Pre-Compression

Yuli Starosvetsky, M. Arif Hasan and Alexander F. Vakakis

*Department of Mechanical Science and Engineering,*
*University of Illinois at Urbana–Champaign,*
*1206 West Green Street, Urbana, IL 61822, USA*
staryuli@illinois.edu, mhasan5@illinois.edu, avakakis@illinois.edu

**Abstract**

We report on the strongly nonlinear dynamics of an array of weakly coupled, non-compressed, parallel granular chains subject to a local initial impulse. The motion of the granules in each chain is constrained to be in one direction which coincides with the orientation of the chain. We show that in spite of the fact that the applied impulse is applied to a one of the granular chains, the resulting pulse that initially propagating only in the excited chain gets gradually equi-partitioned between its neighboring chains, and eventually in all chains of the array. In particular, the initially strongly localized state of energy distribution evolves towards a final stationary state of formation of identical solitary waves that propagate in each one of the chains. These solitary waves are synchronized and have identical speeds. We show that the phenomenon of primary pulse equi-partition between the weakly coupled granular chains can be fully reproduced in coupled binary models which constitute a significantly simpler model that captures the main qualitative features of the dynamics of the granular array. The results reported herein are of major practical significance, since it indicates that the weakly coupled array of granular chains is a medium in which an initially localized excitation gets gradually defocused, resulting in drastic reduction of propagating pulses as they are equi-partitioned among all chains.

## 1. Introduction

The dynamics of one-dimensional homogeneous granular chains excited by various types of loads such as impacts, periodically applied forces (e.g., harmonic loads, periodic impulses etc.) is an area of intensive recent research. It is well known that ordered granular media can be effective attenuators of propagating pulses, so they can be applied effectively as dissipaters of applied shocks. For example, granular layers from iron shot have been used for dampening shocks in explosive chambers, and were shown to be effective mitigators of shock waves formed at the chamber walls.

Granular materials constitute a class of highly nonlinear media where commonly considered linear and weakly nonlinear methodologies are far from being valid. This led to the observation by Nesterenko [1] that an uncompressed homogeneous granular chain possesses zero speed of sound and completely lacks any linear acoustics, leading to its characterization as *sonic vacuum*. This strongly nonlinear dynamics is caused by the essentially nonlinear (non-linearizable) Hertzian interactions between the elastic beads of the chain under compression, and, by the separations between beads and ensuing bead collisions in the absence of compressive loads. Indeed, depending on the applied pre-compression, the dynamics of ordered granular media can range from being strongly nonlinear and highly discontinuous (in the absence of pre-compression) to weakly nonlinear and smooth (in the case of strong pre-compression). Nesterenko and co-workers [1–4] were the first to show that perfectly aligned, one dimensional granular chains can support a special strongly localized traveling wave, the so-called Nesterenko solitary wave. In two additional works general existence theorems for solitary waves in nonlinear [6] and granular [7] chains were proved, based on a general mathematical theorem restricted to one-dimensional nonlinear lattices given by Friesecke and Wattis [8]. In additional works it was shown that despite the complete lack of linear acoustics, uncompressed granular media possess very rich nonlinear dynamics, such as families of travelling waves [9], nonlinear normal modes [10], propagating and standing breathers [11], and countable infinities of resonances and anti-resonances [12].

In this work we consider weakly coupled arrays of homogeneous granular chains (i.e., perfectly ordered granular chains composed of identical elastic spherical beads), and show that pulses that initially are localized to a single chain, gradually 'disperse' to all chains, so that ultimately pulse equipartition occurs. Due to equipartition the initial local pulse gives rise to a set of synchronous solitary waves, albeit of drastically lower amplitudes, each propagating with the same speed at each of the granular chains of the weakly array. To get an understanding of this dynamics we consider a simplified binary collision model which fully captures the main qualitative features of pulse equipartition.

## 2. Pulse equipartition in the granular array

We start our discussion with the formulation of a dynamical system comprising of the array of $N$ weakly coupled, parallel un-compressed homogeneous granular chains. Each granular chain is composed on identical elastic spherical beads that, under compression, interact through an essentially nonlinear (non-linearizable) Hertzian law; in the absence of compressive forces separations between beads are possible leading to possible collisions between beads. Hence, each un-compressed granular chain is a highly nonlinear and possibly discontinuous dynamical system. We assume that an initial impulsive excitation is applied to one of the chains of the array, and we wish to study how the initial impulsive energy is 'dispersed' through the granular array. The governing equations of motion of the array are given by,

$$\ddot{x}_i^{(1)} = \{x_{i-1}^{(1)} - x_i^{(1)}\}_+^{3/2} - \{x_i^{(1)} - x_{i+1}^{(1)}\}_+^{3/2} - \varepsilon\alpha\{x_i^{(1)} - x_i^{(2)}\}$$

$$\ddot{x}_i^{(2)} = \{x_{i-1}^{(2)} - x_i^{(2)}\}_+^{3/2} - \{x_i^{(2)} - x_{i+1}^{(2)}\}_+^{3/2} - \varepsilon\alpha\{x_i^{(2)} - x_i^{(1)}\} - \varepsilon\alpha\{x_i^{(2)} - x_i^{(3)}\}$$

...  (1)

$$\ddot{x}_i^{(N)} = \{x_{i-1}^{(N)} - x_i^{(N)}\}_+^{3/2} - \{x_i^{(N)} - x_{i+1}^{(N)}\}_+^{3/2} - \varepsilon\alpha\{x_i^{(N)} - x_i^{(N-1)}\}$$

$$\dot{x}_1^{(1)}(0) = V, \ x_1^{(1)}(0) = 0, \ \dot{x}_p^{(q)}(0) = 0, \ x_p^{(q)}(0) = 0, \ p, q \neq 1$$

$$i = 1, 2, ...$$

where $x_i^{(j)}$ denotes the displacement of the $i$-th bead in the $j$-th chain, $i = 1, 2, ..., \ 1 \leq j \leq N$. The subscript (+) in a curly bracket indicates that the fractional power expression has meaning only when the argument inside the bracket is non-negative, and otherwise should be set equal to zero. This models the possibility of separations between beads in the absence of compressive forces. In addition, the strength of the lateral coupling between chains is governed by the coefficient $\varepsilon\alpha$, where $\alpha$ is a real constant and $\varepsilon$ a small parameter, i.e., $0 < \varepsilon \ll 1$.

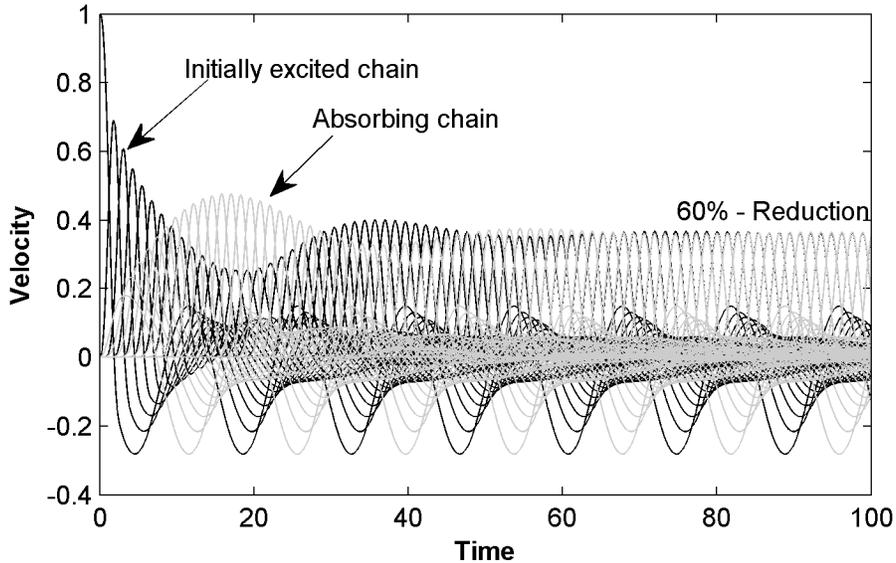

Figure 1. Responses of the first 100 beads of the two chains of the array with $N = 2$ and $\alpha = 1, \varepsilon = 0.1$ for initial excitation of the first chain with an impulse of magnitude $V = 1$.

We start the numerical study of (1) for the case of two, weakly coupled granular chains with $N = 2$ and $\alpha = 1, \varepsilon = 0.1$. Applying an initial impulse of magnitude $V = 1$ to the first particle of the first (excited) chain, we note a quite interesting dynamical phenomenon (cf. Figure 1). Indeed, no matter what is the value of linear lateral coupling $\varepsilon\alpha$ between chains, pulse equipartition between the two granular chains (both excited and absorbing) occurs. That is, after some initial transients during which there is energy exchange between chains in what resembles a nonlinear beat, two identical solitary waves (pulses) are formed in the two chains, being fully synchronized (i.e., with zero phase difference between them), and possessing identical speeds and amplitudes. Hence, a pulse initially localized in one of the chains gets equipartitioned between the two chains. The pulses that eventually form in each of the chains have the form of Nesterenko solitary waves [1] with amplitudes equal to nearly 40% of the initial localized impulse. This is deduced from the results shown in Figure 2, where the responses of the $55^{th}$ beads of the directly excited and absorbing chains are depicted. We note there is formation of identical and synchronized solitary waves in both chains (obviously propagating with identical speeds), so that the weak links between the two chains have no effect on the dynamics after pulse gets equipartion and the formation of the solitary waves in the stationary regime of the dynamics. As a result of pulse equipartition we note a drastic reduction (nearly 60%) of the formed solitary waves in the stationary regime.

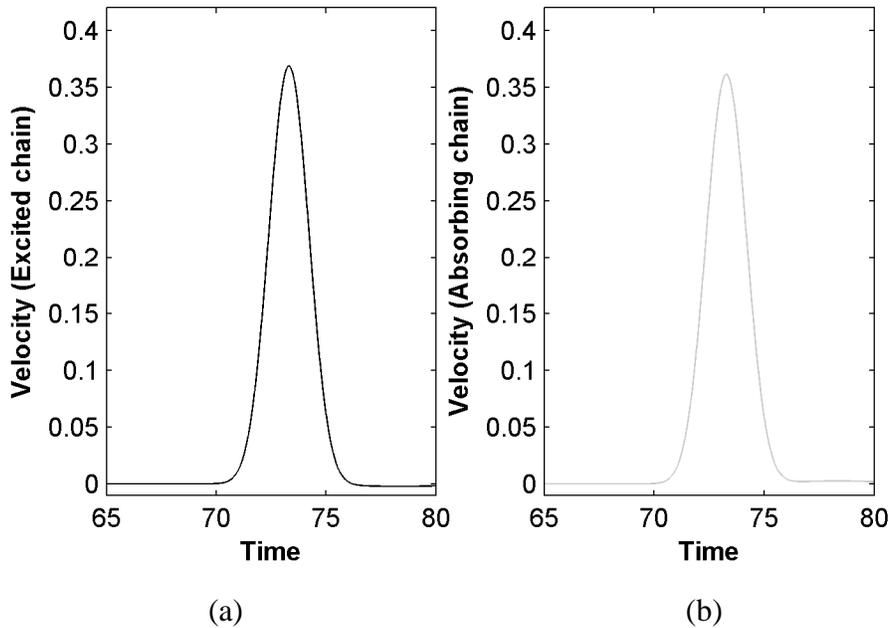

(a)  (b)

Figure 2. Response of the 55th bead of the directly excited (a) and absorbing chain (b) depicting the formation of Nesterenko solitary waves due to equipartition, for the system with $N = 2, V = 1, \alpha = 1, \varepsilon = 0.1$.

Moreover, it appears that pulse equipartition is insensitive to variation of the weak coupling between chains. This is deduced from the plots of Figures 3 and 4, where pulse equipartition is shown for an array with the same initial impulsive excitation, $V = 1$, but with stronger lateral coupling, $\alpha = 5, \varepsilon = 0.1$. In this case, however, the period of initial transients is significantly reduced, and the eventual reduction of the amplitudes of the two formed Nesterenko solitary waves at the stationary regime is nearly 65% compared to the magnitude of the initial localized impulse. This indicates that the phenomenon of pulse equipartition in the weakly coupled array is robust to changes in weak lateral coupling, and more importantly, that the amplitudes of the solitary waves in the two chains that result due to equipartition are dependent on the weak lateral coupling.

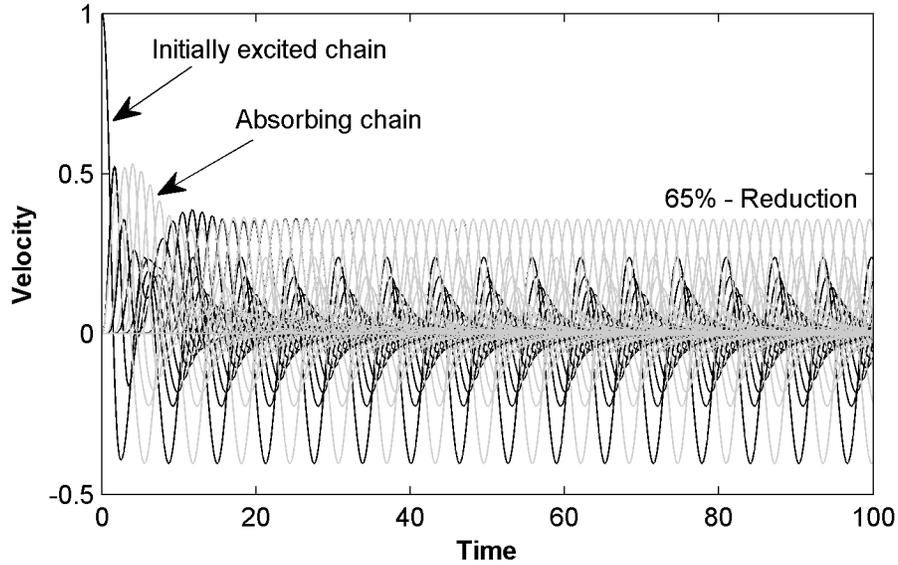

Figure 3. Responses of the first 100 beads of the two chains of the array with $N = 2$ and $\alpha = 5, \varepsilon = 0.1$ for initial excitation of the first chain with an impulse of magnitude $V = 1$.

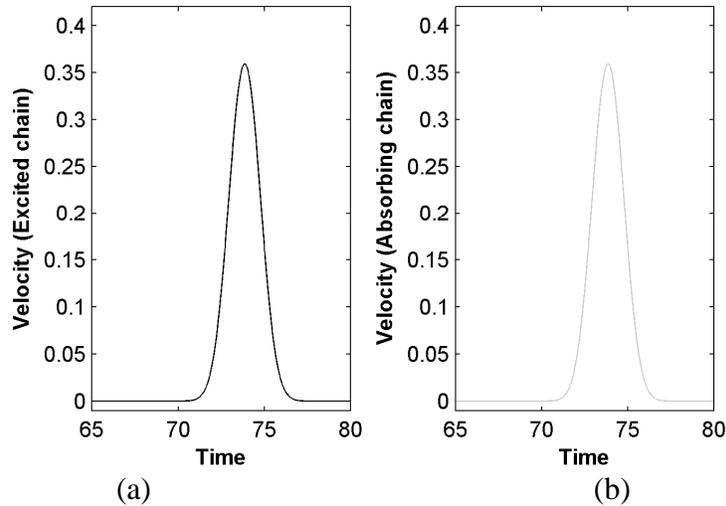

Figure 2. Response of the 55th bead of the directly excited (a) and absorbing chain (b) depicting the formation of Nesterenko solitary waves due to equipartition, for the system with $N = 2, V = 1, \alpha = 5, \varepsilon = 0.1$.

In the next series of simulations we demonstrate that pulse equipartition is robust to the number of parallel chains $N$ in the weakly coupled array. The response of the array with $N = 3$ coupled homogeneous granular chains is depicted in Figure 5 for $V = 1, \alpha = 1$ and $\varepsilon = 0.1$. It is concluded from these results, that in similarity to the array with two chains, there is an initial exchange of energy between the three chains, after which the dynamics settles in a stationary regime where three synchronized Nesterenko solitary waves are formed, one in each of the three parallel chains. Again, in the stationary regime, due to synchronization of three solitary waves are insensitive to the weak lateral links between chains. Moreover, it is clear from these results, that after the energy of the initial impulse is equipartitioned between the three chains, there is a drastic reduction of the amplitudes of the synchronized solitary waves by approximately 75% of the initially applied localized impulse. In Figure 6 we depict the three synchronized identical solitary waves formed in the system in the stationary regime.

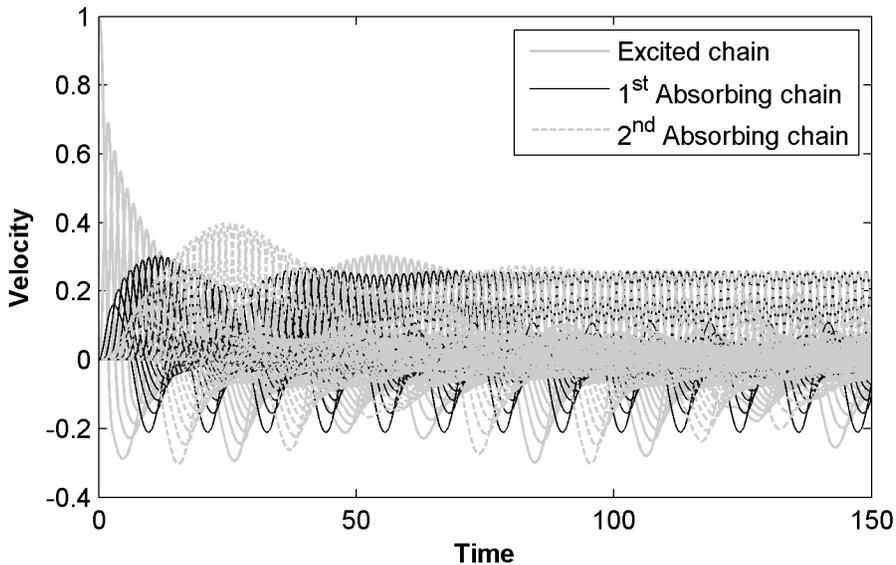

Figure 5. Responses of the first 100 beads of the three chains of the array with $N = 3$ and $\alpha = 1, \varepsilon = 0.1$ for initial excitation of first chain with an impulse of magnitude $V = 1$.

### 3. Modeling by binary collision model

The previous results demonstrate an initial localized energy provided to the first chain of the weakly coupled array is eventually equipartitioned among all chains, and and there is a drastic reduction of the synchronized Nesterenko solitary waves that are eventually formed in the chains. After the formation of the solitary waves the dynamics becomes insensitive to weak later coupling. In order to gain an understanding of the nonlinear mechanism of equipartition we resort to the Binary Collision model first introduced by Rosas and Lindenberg [13], which, as it will be shown, undergoes similar equipartition in its dynamics, and due to its simplicity (compared to the original granular array) is more amenable to theoretical study.

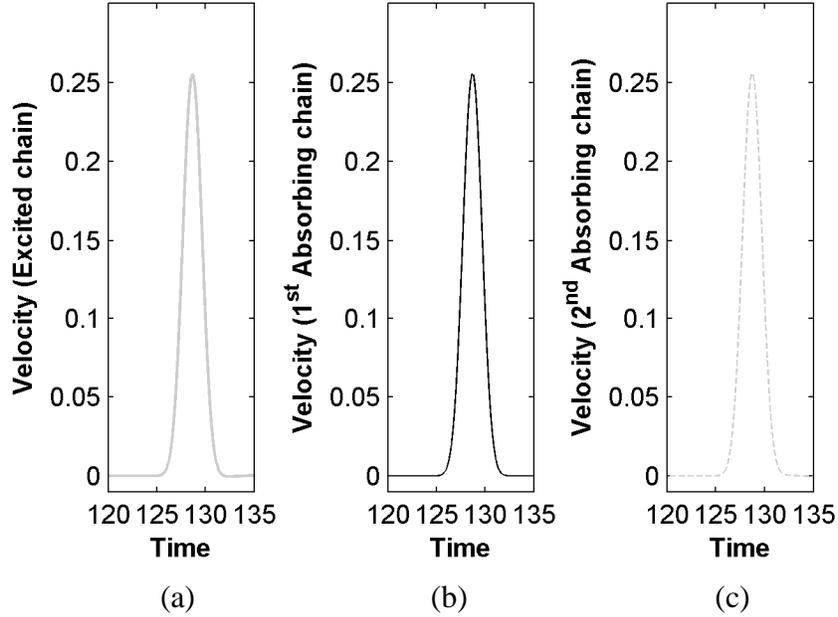

Figure 6. Response of the 90$^{th}$ bead of the directly excited (a), 1$^{st}$ absorbing (b), and 2$^{nd}$ absorbing chain (c), depicting the formation of Nesterenko solitary waves due to equipartition, for the system with $N = 3, V = 1, \alpha = 1, \varepsilon = 0.1$.

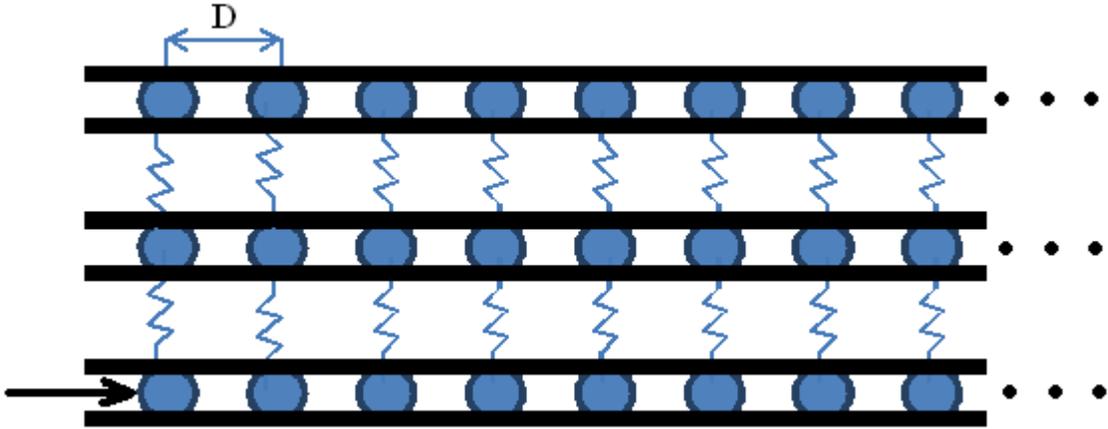

Figure 7. Binary Collision model for studying pulse equipartition.

Hence, we formulate a similar problem to the array of weakly coupled chains, by considering a set of parallel chains with each chain composed of equidistantly placed rigid particles (at constant distance $D$) that collide elastically with each other. This binary collision model is depicted in Figure 7; its main advantage is that there can be only collisions between two particles at any given time (since the collisions are instantaneous), which significantly simplifies the dynamics. We note that in the absence of coupling the dynamics of each chain of the binary model is integrable and analyzable. By contrast, in the granular array (1) the durations of elastic collisions between beads is finite, so that simultaneous between more than two beads can be realized; this renders the dynamics non-integrable, and difficult to analyze. Considering that an impulsive excitation is applied locally to one of the chains of the binary collision model, we study the 'dispersion' of the initial pulse through the array of coupled particles chains.

Examining the response of the binary model we note a quite similar equipartition phenomenon to what has been observed for the array of granular chains (1). In Figure 8 we consider a binary model with $N = 2$ chains of particles, a unit impulsive excitation and coupling parameters $\alpha = 1, \varepsilon = 0.1$, and depict the transient response of the fifth bead in each of the two chains, as well as the residual oscillations or 'oscillating tail' following the transient response; this 'oscillating tail' arises due to the weak lateral coupling between chains. As time progresses, however, and the pulse propagates downstream through the array, a stationary state of the dynamics is reached and the primary pulse reaching the $25^{th}$ bead of each of the chains of the binary model contains nearly zero oscillations and has the form of a solitary wave (cf. Figure 9). Moreover, the solitary waves in the two chains are synchronized, and of identical speeds and amplitudes. It follows that the dynamics observed in the binary model chain is closely related to the dynamics if the array of granular chains (1).

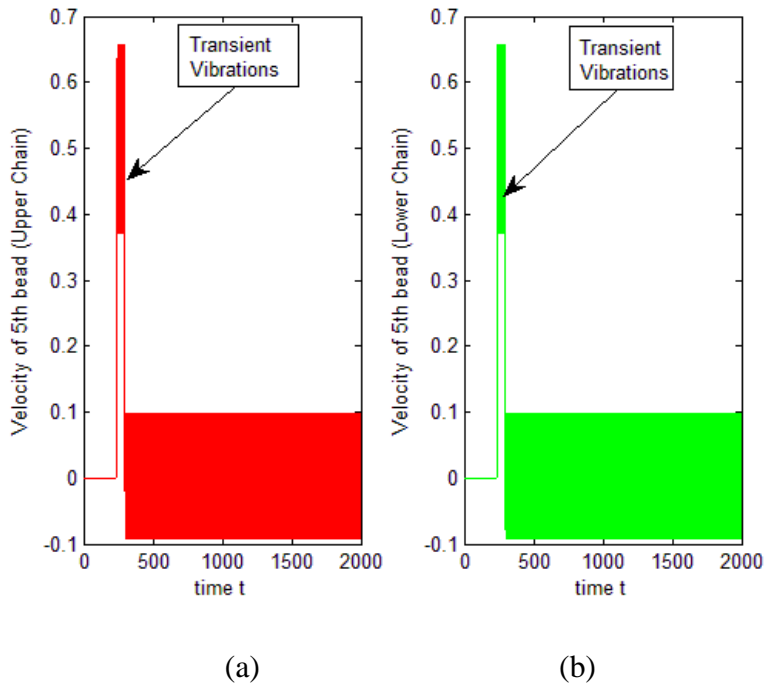

Figure 8. Transient response of the $5^{th}$ bead of (a) the excited chain, and (b) the absorbing chain of the binary model before the stationary state and pulse equipartition are reached.

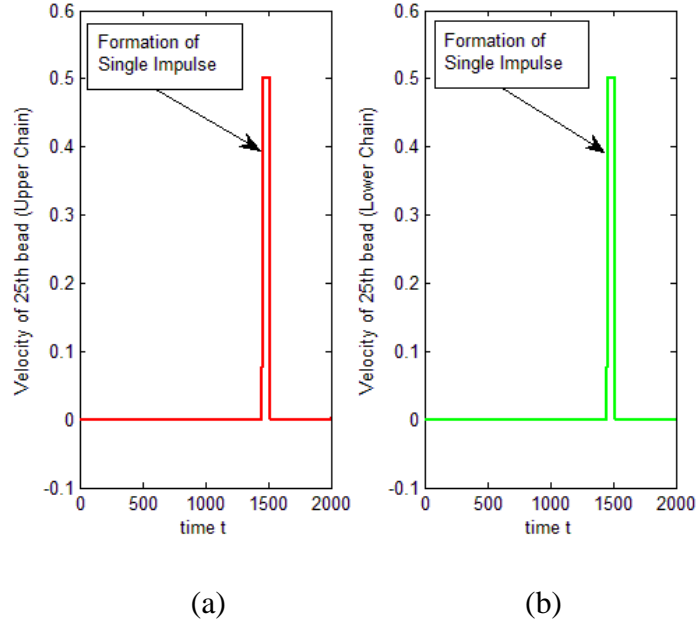

(a)                  (b)

Figure 9. The stationary state and pulse equipartition in the Binary Collision model: Response of the 25$^{th}$ bead of (a) the excited chain, and (b) the absorbing chain.

As we can see from the simulation results, this simplified binary collision model may successfully capture the mechanism of equipartition and can provide good estimates the solitary waves formed in the stationary state of the dynamics. At this point it will be instructive to examine the possible case of formation of a stationary state of equi-partition in the binary model. To this end we consider the system of well separated granules in two chains that are coupled by means of weak linear springs. In the present discussion we refer to each pair of coupled granules as 'dumbbell'. Such a structure can be described by the following system of differential equations,

$$\ddot{x}_i^{(1)} = \left(x_{i-1}^{(1)} - x_i^{(1)} - D\right)_+^{3/2} - \left(x_i^{(1)} - x_{i+1}^{(1)} - D\right)_+^{3/2} + \alpha\left(x_i^{(2)} - x_i^{(1)}\right)$$
$$\ddot{x}_i^{(2)} = \left(x_{i-1}^{(2)} - x_i^{(2)} - D\right)_+^{3/2} - \left(x_i^{(2)} - x_{i+1}^{(2)} - D\right)_+^{3/2} + \alpha\left(x_i^{(1)} - x_i^{(2)}\right)$$
$$x_i^{(1)}(0) = 0, \dot{x}_i^{(1)}(0) = 0, x_i^{(2)}(0) = 0, \dot{x}_i^{(2)}(0) = 0, \quad i = 1, 2, ...$$
$$x_1^{(1)}(0) = 0, \dot{x}_1^{(1)}(0) = 1, x_1^{(2)}(0) = 0, \dot{x}_1^{(2)}(0) = 0$$
(2)

where the new parameter $D$ denotes the (fixed) distance between neighboring beads in each chain. Therefore the Hertzian contact forces are not active as long as the beads do not touch one another in compression. It can be demonstrated analytically (by transforming system (2) into coordinates of the center of mass of each of the dumbbells) that the initially applied impulse to the first dumbbell will produce a primary pulse that will propagate through the chain undisturbed. This is also apparent from simple arguments based on conservation of linear momentum. However, the main question to be asked concerns the partition of the primary pulse among the two weakly coupled chains. To answer this question in a complete rigorous way, even for the case of binary collisions model is a formidable task, however in the present discussion we can provide a particular and clear example for the possibility for equi-partition of propagating pulses in the binary model.

To this end, we approximately analyze the dynamics of (2) by transforming to coordinates of the center of mass of each chain and to relative displacements between neighboring beads. Considering the motion of the first dumbbell from the initial time when the impulse is applied and before the resulting primary pulse propagates to the second dumbbell, the motion of the first bead of each chain is approximately governed by:

$$\ddot{x}_1^{(1)} \approx \alpha \left( x_1^{(2)} - x_1^{(1)} \right)$$
$$\ddot{x}_1^{(2)} \approx \alpha \left( x_1^{(1)} - x_1^{(2)} \right) \qquad (3)$$
$$x_1^{(1)}(0) = 0, \dot{x}_1^{(1)}(0) = 1, x_1^{(2)}(0) = 0, \dot{x}_1^{(2)}(0) = 0$$

This yields the solution:

$$u_1(t) = \frac{1}{2}\left( t + \frac{1}{\sqrt{2\alpha}} \sin\left(\sqrt{2\alpha}t\right) \right), \quad \dot{u}_1(t) = \frac{1}{2}\left(1 + \cos\left(\sqrt{2\alpha}t\right)\right)$$
$$u_2(t) = \frac{1}{2}\left( t - \frac{1}{\sqrt{2\alpha}} \sin\left(\sqrt{2\alpha}t\right) \right), \quad \dot{u}_2(t) = \frac{1}{2}\left(1 - \cos\left(\sqrt{2\alpha}t\right)\right) \qquad (4)$$

Now we can select the parameter *D* in (2) in such a way that the second collision will occur exactly at a time when the velocities of the two leading beads in the two chains are equal, yielding the relation:

$$D = \frac{\pi + 2}{4\sqrt{2\alpha}} \qquad (5)$$

In this case the first collision will occur simultaneously for both beads of the dumbbell with equal velocities, which will finally result in the independent propagation of identical impulses on both chains. This is only a simple example for achieving the equi-partition mechanism for an initially localized impulse. It should be also noted that similar route to equi-partition of pulses has been also observed numerically for rather different values of the parameter *D*, however the complete rigorous proof of equi-partition for an arbitrary value of *D* is a challenging task that will not pursued here. In fact the consideration of binary collision models can reveal the very important analogy between the propagating solitary waves (compactons) in the full granular chain and the propagating impulses in the binary models which exhibit a similar phenomenon and also exhibit their own routes to equi-partition. The very important task to be considered in the future is to find the necessary conditions for the binary chain for equi-partition and based on these to proceed to the parametric study of the full granular system (1).

## 4. Concluding Remarks

We report a very interesting phenomenon observed in the dynamics of an array of weakly coupled, uncompressed granular chains subject to localized excitations. Namely, we demonstrate numerically that following the localized excitation of one of the chains of the array there occurs gradual leakage of energy to all neighboring chains in a process that initially appears in the form of nonlinear beating, but eventually reaching a stationary state wherein equipartition of energy to all chains occurs, signified by the formation of synchronous Nesterenko solitary waves with identical speeds and amplitudes in each of the chains of the array. Hence, a localized excitation eventually 'spreads' evenly among all chains; this pulse equipartition is robust to variations of the uniform lateral coupling between chains, as long as it is weak. In addition, we have shown that a simplified binary model exhibits a similar phenomenon of pulse equipartition, so that this interesting nonlinear dynamical mechanism can be understood and explained by this binary model which can be studied analytically. These results are of significant practical importance in engineering applications dealing with passive shock wave attenuation, since it is quite evident that a material system incorporating the type of granular setups discussed herein carries the capacity to defocus and effectively disperse initially localized applied pulses.

**Acknowledgments**

This work was funded by MURI grant US ARO W911NF-09-1-0436, Dr. David Stepp is the grant monitor.